\RequirePackage{fix-cm}
\documentclass[final,authoryear,3p,12pt,times]{svjour3}                     % onecolumn (standard format)
\smartqed  % flush right qed marks, e.g. at end of proof
\usepackage{graphicx}
\graphicspath{{plots/}}

\journalname{Environmental Fluid Mechanics}
\usepackage{epsfig,bm,color}
\usepackage{lscape}
\usepackage{amssymb,amsmath}
\usepackage{array,booktabs,tabularx}
\usepackage{amssymb}
\usepackage{bm}
\newcolumntype{Z}{>{\centering\arraybackslash}X}
\usepackage[]{units}
\usepackage{mathtools}

\usepackage{times}
\usepackage{epstopdf}
\usepackage{caption}
\usepackage{amsfonts}
\usepackage[T1]{fontenc}
\usepackage[table]{xcolor}
\usepackage{booktabs}
\usepackage{multirow}
\usepackage{multicol}
\usepackage{longtable}
\usepackage{mhchem}
\usepackage{listings}

\usepackage{fancyvrb}

\begin{document}

\title{Subgrid-scale energy transfer and associated coherent structures in turbulent flow over a forest-like canopy}

\author{Md. Abdus Samad Bhuiyan         \and
        Jahrul M Alam %etc.
}


\institute{Md Abdus Samad Bhuiyan  \at
  Department of Mathematics and Statistics, Memorial University of Newfoundland, St.John's, NL, Canada\\               
  \email{masb80@mun.ca,} \\          %  
  \emph{Present address:} Software Developer, MKIT North America Inc, Canada %  if needed
  \and
  Jahrul M Alam \at
  Department of Mathematics and Statistics, Memorial University of Newfoundland, St.John's, NL, Canada
}

\date{}

\maketitle

\begin{abstract}
  Large eddy simulation allows to incorporate the important driving physics of turbulent flow through forest- or vegetation-like canopies. In this paper we investigate the effects of  vortex stretching and coherent structures on the subgrid-scale (SGS) turbulence kinetic energy (TKE). We present three simulations (SGS-d/s/w) of turbulence-canopy interactions. SGS-d assumes a local, dynamic balance of SGS production with SGS dissipation. SGS-s averages the SGS contributions of coherent structures over Lagrangian pathlines. SGS-w assumes that an average cascade of TKE from large- to small-scales occurs through the process of vortex stretching. We compare the consequences of considering a forest of the same morphology as immersed solids or an immersed canopy. Our results show clear differences in the characteristics of flow and turbulence, while both the cases exhibit canopy mixing layers. The results also show that the consideration of vortex stretching resolves about $18$\% more TKE with respect to classical Deardorff's TKE model. These observations indicate that the aerodynamic response of the forest canopy is linked to the morphology of the forest cover. Sweep- and ejection-events of the spatially intermittent coherent structures in forests as well as their role in transporting momentum, energy, and scalars are discussed. %

\end{abstract}


\section{Introduction}
In the atmospheric boundary layer~(ABL), the aggregate effect of a large collection of vertical obstacles (e.g., trees, buildings, etc)~\cite{Brunet94,Stvredova2012,Philips2013,Park2015,Miri2017} is to squeeze the turbulence-containing surface layer between the outer layer and the canopy-roughness sublayer~\cite{Shaw92,Finnigan2000,Finnigan2009} . For example, the Earth Observatory Report of NASA indicates the hight ($h$) of forests can be up to $70$~m. The roughness effect of such a forest may influence a height $3h$ to $5h$ from the ground~\cite{Belcher2012}. During the past three decades, the turbulence kinetic energy~(TKE) budget of such canopies was thoroughly scrutinized by wind-tunnel and large-eddy simulation (LES). Past work suggest that the pressure transport is dominant in the lowest two-third ($2h/3$) of forest canopies~\cite{Brunet94,Finnigan2000,Marjoribanks2016}; however, large-scale coherent structures are primarily responsible for the turbulent transfer between forests and the atmosphere~\cite{Shaw92,Raupach96,Dupont2009,Finnigan2009,Konstantin2018}. A substantial work also focused on parameterizing the aggregate effects of the form drag induced by the canopies. There are few important questions associated with the transition of length scales of coherent structures, such as how to realistically represent the unresolved part of the turbulent transport in the canopy-roughness sublayer, where the energy-containing motions are inherently under-resolved by LES. 

In this article we follow the vorticity transport theory of Taylor~\cite{Taylor32} in which subgrid scale turbulence is regarded as an effect of vortex stretching rather than as a diffusion of straining motion. The local differences in pressure would affect the momentum of an eddy, and thus, the strain rate provides incorrect rate of turbulence dissipation, thereby requiring {\em ad-hoc} modification of SGS models for ABL flows. Based on the vortex stretching hypothesis, we present LES results for a turbulent flow over forest-like canopies in ABL. Vortex stretching appears as an appealing candidate for SGS modeling because the direction of maximal stretching is the direction of maximum positive eigenvalues of the strain tensor. In other words the stretching of vortices sets the rate of dissipation through the enstrophy production. The consideration of vortex stretching can be useful for canopy flows because in the surface layer, the energy-containing length-scale diminishes, and anisotropy strengthens. The most widely used SGS models for ABL flows~\cite{Moeng2015} needs dynamic adjustments of model parameters in order to capture the TKE budget and the local backscatter of energy. In LES ({\em e.g.}~\cite{Deardorff70,Deardorff73,Deardorff74,Deardorff80}), vortex stretching could be directly linked to the energy containing resolved modes of the solution of Navier-Stokes equations. Recent work on LES of canopy flows indicate that vortex stretching and coherent structures can be useful for advancing subgrid models~\cite{Shaw92,Finnigan2000,Dupont2008,Dupont2009,Finnigan2009,Yan2017,Konstantin2018}. Some investigations~({\em e.g.} \cite{Finnigan2000,Finnigan2009,Dupont2009}) have emphasized that the morphology of the canopy usually leads to considerable scatter of the SGS coherent motion in the roughness sublayer, which is dominated by the stretching of small-scale vortices. 

In LES, the turbulence eddy-viscosity
\begin{equation}
  \label{eq:nu}
   \nu_{\hbox{\tiny sgs}} = (C_s\Delta_{\hbox{\tiny LES}})^2(2\mathcal S_{ij}\mathcal S_{ij})^{1/2}
\end{equation}
was proposed by Smagorinsky~\cite{smagorinsky}, which ensures that $-\mathcal S_{ij}\tau_{ij} + C_{\epsilon} k^{3/2}_{\hbox{\tiny sgs}}/\Delta_{\hbox{\tiny LES}} = 0$; {\em i.e.} the dissipation $-C_{\epsilon} k^{3/2}_{\hbox{\tiny sgs}}/\Delta_{\hbox{\tiny LES}}$ accounts for the local production $-\mathcal S_{ij}\tau_{ij}$ if $\tau_{ij} - (3/2) k_{\hbox{\tiny sgs}}\delta_{ij} = 2C_k\Delta_{\hbox{\tiny LES}}k^{1/2}_{\hbox{\tiny sgs}}\mathcal S_{ij}$ and $C_s^2=C_k^{3/2}/C_{\epsilon}$ (see~\cite{Moeng2015}). Here, $\tau_{ij}$ and $\mathcal S_{ij}$ are subgrid-scale stress and resolved strain, respectively. Deardorff~\cite{Deardorff70} considered a local dynamic balance of the energy flux $-\tau_{ij}\mathcal S_{ij}$  that is to be transferred from the resolved scale motion and dissipated by the unresolved motion. This approach solves a transport equation in order to provide the eddy viscosity $\nu_{\hbox{\tiny sgs}} = C_s\Delta_{\hbox{\tiny LES}}k^{1/2}_{\hbox{\tiny sgs}}$~\cite{Deardorff70}. Meneveau {\em et. al.}~\cite{Meneveau96} proposed to dynamically calculate $C_s$ by averaging the Lagrangian history of coherent structures, which is very useful when the grid is not sufficient to resolve the canopy-roughness sublayer. Sullivan et al~\cite{Sullivan94} and Leveque~\cite{Leveque2007} proposed to locally adjust $\nu_{\hbox{\tiny sgs}}$ by considering the fluctuating strain rates in Eq~(\ref{eq:nu}) instead of the resolved one. Instead of considering the rate of strain $\mathcal S_{ij}$, Nicoud~\cite{Nicoud99} considered the trace-less symmetric part of the square of the velocity gradient tensor for estimating $\nu_{\hbox{\tiny sgs}}$. In canopy flows, it is important to consider that the energy cascade from large to small scales depends on two different mechanisms, namely, a linear transfer that accounts for the mean rate of strain $\mathcal S_{ij}$ in a statistical sense and a nonlinear transfer that accounts for the transition of scales of coherent structures. The guideline for such a consideration in the present study of canopy flows is a continuation of the work of~\cite{Sullivan94}, \cite{Nicoud99}, and \cite{Leveque2007}.

In what follows, we begin by a general outline of canopy layer, porous media, and turbulence modeling in Sec~\ref{sec:meth}. We then consider a brief comparison of the result of proposed LES with the wind-tunnel measurements in Sec~\ref{sec:comp}, before highlighting the potential differences in dynamic modeling of SGS turbulence in Sec~\ref{sec:and}. 

\section{Materials and methods}\label{sec:meth}
\subsection{Aerodynamic response to forest morphology and theory of porous media}
The method of volume average is applied to the Navier-Stokes equation~(NSE) to model the fluid dynamic response to solid obstacles as a drag force $f_i$ that accounts for the pressure and the viscous stress~\cite{Finnigan2000,Finnigan2009}. If a box-filter is applied to NSE, we have the sub-filter scale kinematic stress $\tau_{ij}=\langle \tilde u_i\tilde u_j\rangle - \langle \tilde u_i\rangle\langle \tilde u_j\rangle$, where $\tilde u_i = \langle u_i\rangle + u'_i$ is the total (filtered+subgrid) velocity in the $x_i$ direction. The volume averaging and the box-filtering commutes except in the canopy region. For simplicity, we may drop the symbol `$\langle\cdot\rangle$' from the filtered variable, unless it is explicitly needed.

The finite volume method acts as an implicit box-filtering kernel, which separates the motion beyond the cutoff scale $\Delta_{\hbox{\tiny LES}} (\propto\sqrt[3]{\Delta x\Delta y\Delta z})$, where $(x_1,x_2,x_3) = (x,y,z)$. In this study, we solve the filtered NSE for the atmospheric boundary layer flow, where the aerodynamic response to forest morphology extends from the surface $x_3=0$ to a height of $x_3=h$.  The filtered equations are (where the summation convention is assumed in the rest of the article)
\begin{equation}\label{eq:LES1}
 \frac{\partial{u_i}}{\partial{t}}+u_j\frac{\partial{ u_i}}{\partial{x_j}}=-\frac{1}{\rho_0}\frac{\partial{\langle P\rangle}}{\partial{x_i}} - \frac{\partial p}{\partial x_i} - \frac{\partial{\tau_{ij}}}{\partial{x_j}} - f_i, %\frac{\partial{\tau^c_{ij}}}{\partial{x_j}},
\end{equation}
\begin{equation}\label{eq:LES2}
\hbox{and }\quad\frac{\partial{u_i}}{\partial{x_i}} = 0.
\end{equation}
In the literature~\cite{Shaw92}, a canonical framework of representing forest or vegetation canopies is to replace $f_i$ in the momentum equation~(\ref{eq:LES1}) by a pressure-drag~\cite{Dupont2008}. In the classical theory of porous media, $f_i$, in Eq~(\ref{eq:LES1}), is expressed through the Darcy-Forchheimer model of  a porous layer~\cite{DeLemos2006} in which a portion of the pressure gradient of Eq~(\ref{eq:LES1}) accounts for
\begin{equation}
  \label{eq:DFM}
  \frac{\partial p''}{\partial x_i} = -\left(\frac{\nu}{K} + \frac{C_d|\bm u|}{\sqrt{K}}\right)u_i,
\end{equation}
where $p''$ is the local pressure change experienced by the canopy, $K$ is the porosity, and $C_d$ is a constant. It was reported in~\cite{Finnigan2009} that the volume averaged pressure drag -- the last term in~(\ref{eq:DFM}) -- is about three times larger than the viscous drag -- the second last term in~(\ref{eq:DFM}). The kinematic pressure-drag of a canopy is typically proportional to the product of a one-sided plant area density $\mathcal A\sim K^{-1/2}$ and the square of the resolved velocity~\cite{Dwyer97}. %We have retained the viscous term in Eq~(\ref{eq:DFM}). %For all simulations reported in this work, the canopy drag force $f_i$ in~(\ref{eq:LES1}) is expressed as $f_i = C_d\mathcal A|\bm u|u_i$, where $\bm u = \sqrt{u_iu_i}$.

\subsection{Turbulence modeling}
As discussed in the introduction, turbulent motions at scales smaller than the grid spacing are accounted for through the deviatoric subgrid stress $\tau_{ij}^{\hbox{\tiny sgs}} = \tau_{ij}-(1/3)\tau_{kk}\delta_{ij}$ in the second last term of Eq~(\ref{eq:LES1}), which is computed using the strain rate as
\begin{equation}
  \label{eq:sgs}
  \tau_{ij}^{\hbox{\tiny sgs}} = \nu_{\hbox{\tiny sgs}}\mathcal S_{ij}.
\end{equation}
In~(\ref{eq:sgs}), it is assumed that the local pressure differences have no effect on the momentum transported by large eddies. Thus, the eddy viscosity $\nu_{\hbox{\tiny sgs}}$ is defined to be a product of a length scale $\Delta_{\hbox{\tiny LES}}$ and a velocity scale $\Delta_{\hbox{\tiny LES}}(2\mathcal S_{ij}\mathcal S_{ij})^{1/2}$, where the Smagorinsky-Lilly parameter $C_s$ determines the desired contribution of unresolved eddies or coherent structures. As the surface is approached, anisotropy in turbulence increases, and thus, the dynamic approach~\cite{Germano91} of evaluating $C_s$ directly from the resolved field becomes important.  %Canopy flows share features of mixing layer and boundary layer~\cite[see][]{Finnigan2009}, which suggests that the magnitude of $\mathcal S_{ij}$ may overestimate the time scale of subgrid scale coherent structures. 

\iffalse %JA Apr 8, can we skip the BC
At the bottom boundary, a common approach to model the rough wall is to impose the stress boundary condition through the logarithmic law-of-the wall, {\em i.e.},
%
\begin{equation}
  \label{eq:tauw}
  \langle\tau_w^{\hbox{\tiny LES}}\rangle = - \left[\frac{\kappa}{\ln(z_1/z_0)}\right]^2(\langle u_1^2\rangle + \langle u_2^2\rangle)
\end{equation}
%
where $\kappa$ is the von Karmann constant $(0.41)$ and $z_1$ is the first off-wall grid point.

%theturbulent transferprocessisessentially nonlocal and cannot be described by a local gradient-diffusion relationship.
\fi %JA Apr 8

\iffalse
Eq~(\ref{eq:tauw}) may be compared with the log-law in an average sense, {\em i.e.},
\begin{equation}
  \label{eq:tau-log}
  \langle\tau_w^{\hbox{\tiny log-law}}\rangle = - \left[\frac{\kappa}{\ln(z_1/z_0)}\right]^2(\langle u_1\rangle^2)
\end{equation}
where the mean spanwise velocity component $\langle u_2\rangle = 0$. Due to a rough surface, streamwise velocity near the first off-wall grid points would fluctuate like $\langle u_1^2\rangle > \langle u_1\rangle^2$, and thus $\langle\tau_w^{\hbox{\tiny LES}}\rangle > \langle\tau_w^{\hbox{\tiny log-law}}\rangle$. In LES, the rough-surface model~(\ref{eq:tauw}) would introduce additional velocity deficit near the surface. 

Since the viscous and the turbulent stresses are of similar magnitude in the buffer layer ($5 < z^+ < 15$, $z^+ = zu_*/\nu$), we can assume a continuity of the wind profile between the viscous layer, $u^+=z^+$, and the logarithmic layer, $u^+ = (1/\kappa)\ln(z^+/z_0^+)$. Eq~(\ref{eq:tauw}) can be re-formulated using $u_{\tau}^2 =  \langle u_1^2\rangle + \langle u_2^2\rangle$, {\em i.e.}

\begin{equation}
  \label{eq:tausgs}
  \tau_w(x,y) = \langle\tau_w^{\hbox{\tiny LES}}\rangle \left[\frac{z^+\kappa-\ln(z/z_0)}{\ln(z/z_0)}\right].
\end{equation}
%
\fi

\subsubsection{Lagrangian dynamic SGS model (SGS-s)}
In the Lagrangian dynamic procedure, $C_s$ is determined by averaging inhomogeneous flows over a Lagrangian time scale, where one solves two additional transport equations, namely, for $\mathcal I_{lm}$  and $\mathcal I_{mm}$. Using the Germano identity $\mathcal L_{ij} = \mathcal T_{ij}-\tau_{ij}$, where  the additional stress $\mathcal T_{ij}$ comes from the test-filtering operations at a scale $2\Delta_{\hbox{\tiny LES}}$ (or larger), the Lagrangian approach defines the model parameter $C_s = \sqrt{\mathcal I_{lm}/\mathcal I_{mm}}$. This approach is known to be effective for inhomogeneous flows. Yan {\em et. al.}~\cite{Yan2017} observed some discrepancies in the simulation of canopy flows by the Lagrangian dynamic model, which may be attributed to the lack of appropriate test-filtering kernels because solid bodies immersed in the fluid are resolved by the work of~\cite{Yan2017}. In case of considering the classical canopy model, the Lagrangian dynamic procedure can properly represent the role of subgrid-scale energy-containing motion.

\subsection{Deardorff's TKE model for canopy flows (SGS-d)}

Deardorff~\cite{Deardorff80} advanced the Smagorinsky-Lilly model, Eq~(\ref{eq:nu}), by incorporating a dynamic balance of subgrid-scale production and dissipation, where 
\begin{equation}
  \label{eq:nut-k}
  \nu_{\hbox{\tiny sgs}} = C_s k^{1/2}_{\hbox{\tiny sgs}}\Delta_{\hbox{\tiny LES}}
\end{equation}
is dynamically varied by obtaining $k_{\hbox{\tiny sgs}}^{1/2}$ from the TKE equation
\begin{equation}
  \label{eq:tke}
  \frac{\partial k_{\hbox{\tiny sgs}}}{\partial t} + \frac{\partial u_j k_{\hbox{\tiny sgs}}}{\partial x_j} = -\tau_{ij}\mathcal S_{ij} - C_\epsilon\frac{k^{3/2}_{\hbox{\tiny sgs}}}{\Delta_{\hbox{\tiny LES}}} +  \frac{\partial}{\partial x_j}\left(\nu_{\hbox{\tiny sgs}}\frac{\partial k_{\hbox{\tiny sgs}}}{\partial x_j}\right).
\end{equation}
As mentioned earlier, the Smagorinsky model Eq~(\ref{eq:nu}) can be derived by retaining only the first two terms in the right hand side of Eq~(\ref{eq:tke}). The second last term in Eq~(\ref{eq:tke}) can be written as $-C_\epsilon (\ell/\Delta_{\hbox{\tiny LES}})\epsilon_{\hbox{\tiny sgs}}$,  where $\ell = k_{\hbox{\tiny sgs}}^{3/2}/\epsilon_{\hbox{\tiny sgs}}$ is the integral length scale. Notice the new parameter $C_\epsilon$ in Eq~(\ref{eq:tke}). Commonly used values of the two parameters for ABL flow simulations are $C_s\sim 0.1$, and $C_\epsilon\sim 0.93$~\cite{Moeng2015}.

In the present work, we consider the dynamic variation of the parameters $C_s$ and $C_\epsilon$, which are determined using the similar approach of classical dynamic Smagorinsky model~\cite{Germano91}. Following~\cite{Shaw92} we have also added a sink term in Eq~(\ref{eq:tke}) in order to represent the dissipation of turbulent eddies in the canopy zone~\cite{Dupont2008}. Note that ref~\cite{Finnigan2009} considered a constant value of $C_s$, where they diagnosed $k_{\hbox{\tiny sgs}}$ from the resolved flow instead of solving Eq~(\ref{eq:tke}). In addition to determining the velocity scale $k^{1/2}_{\hbox{\tiny sgs}}$ dynamically, the dynamic variation of $C_s$ and $C_\epsilon$ in the present study also provides useful feedback for considering such a model in turbulence resolving atmospheric boundary layer flow simulations. It is worth mentioning that the SGS-d model with constant values of the model parameters is usually considered in simulations of atmospheric boundary-layer flows~\cite{Moeng84,Chow2019}. %The dynamic procedure~\cite[see][]{Huang2010} utilized in this model addresses the instability associated with the classical dynamic Smagorinsky model. %However, it should be noted that this SGS model requires test-filtering operations to calculate the model coefficients and defining a consistent test-filtering operation in complex geometries remains an active research topic.

\subsubsection{Coherent structure-based scale-adaptive SGS model (SGS-w)}
%
%Although a precise definition of the term `coherent structure' for canopy flows is lacking (and highly debated in general), previous work~\cite[e.g.][]{Dupont2008,Finnigan2009} demonstrate that they consist of the regions of spatially-coherent vortical motions in the canopy zone.  
%Here, we extend one of the recent models of the eddy-viscosity $\nu_{\hbox{\tiny sgs}}$ to simulation canopy flows, which is based on a method of coherent structure detection.
A purpose of considering coherent structures in the dynamic modeling approach~(see~\cite{Taylor32}) is to address cross-scale interaction in turbulent flows by introducing scale-awareness in the eddy viscosity model~\cite{Alam2015,Alam2018}. We have seen from the Lagrangian dynamic approach and Deardorff's TKE-based model that it is important to adjust the velocity scale given by $ k_{\hbox{\tiny sgs}}^{1/2} = \Delta_{\hbox{\tiny LES}}(2\mathcal S_{ij}\mathcal S_{ij})^{1/2}$, which improves the Smagorinsky-Lilly model~(\ref{eq:nu}). This is because the characteristic length scales of coherent structures varies as the flow transition from one sublayer to the other as the surface is approached. An algebraic approach of modeling the SGS TKE from the resolved velocity is to consider the scale-adaptive eddy viscosity~\cite{Nicoud99,Nicoud2011,Trias2015}
\begin{equation}
  \label{eq:Nw}
  %\nu_{\hbox{\tiny sgs}} = C_w^2\Delta_{\hbox{\tiny LES}}^2\frac{(\mathcal I^d_2)^{3/2}}{(\mathcal I_2)^{5/2} + (\mathcal I^d_2)^{5/4}},
  \nu_{\hbox{\tiny sgs}} = C_w\Delta_{\hbox{\tiny LES}}\left[\overbrace{\frac{\Delta_{\hbox{\tiny LES}}(\mathcal S^d_{ij}\mathcal S^d_{ij})^{3/2}}{(\mathcal S_{ij}\mathcal S_{ij})^{5/2} + (\mathcal S^d_{ij}\mathcal S^d_{ij})^{5/4}}}^{k_{\hbox{\tiny w}}^{1/2}}\right].
\end{equation}
In Eq~(\ref{eq:Nw}), $\mathcal S_{ij}^d$ is the trace-less symmetric part of the tensor $(\partial u_i/\partial x_j)^2$. It can be shown that the expression with the square bracket $[\cdot]$ has an asymptotic limit of $\mathcal O([zu_*/\nu]^3)$ as $z\rightarrow 0$, where the mean flow varies only in the wall-normal direction. Thus, in the vicinity of a solid within the viscous sublayer, $\nu_{\hbox{\tiny sgs}}$ defined by Eq~(\ref{eq:Nw}) vanishes. We find that $\mathcal S_{ij}^d\mathcal S_{ij}^d = (2/3)Q^2 + (1/2)|\mathcal S\bm\omega|$, where $Q$ is the second invariant of the velocity gradient tensor and $\mathcal S\bm\omega=\mathcal S_{ij}\bm\omega_j$ is the vortex stretching vector. In other words, if we consider the expression within $[\cdot]$ in Eq~(\ref{eq:Nw}) for computing $k_{\hbox{\tiny sgs}}^{1/2}$, it would provide a velocity scale ({\em i.e.} the eddy viscosity) which is dynamically adjusted according to the magnitude of the coherent structure and vortex stretching. Equating the eddy-viscosity given by Eq~(\ref{eq:Nw}) to that given by Eq~(\ref{eq:nu}), the constant $C_w$ can be determined in terms of the resolved velocity and the Smagorinsky-Lilly constant $C_s$. For the numerical tests reported in this paper, we have found that a constant value of $C_w=0.325$ provides reasonably accurate results in comparison to the two other models considered in the present study.

\section{Comparison with wind-tunnel measurements}\label{sec:comp}
\subsection{Experimental setup}
A wind-tunnel model of canopy was designed with cylindrical stalks of diameter $0.00025$~m and length $0.05$~m~\cite{Brunet94}. The stalks were arranged on a uniform square grid of side $0.005$~m. Briefly, the working section of the wind tunnel was $11$~[m] long, $1.8$~[m] wide, and $0.65$~[m] high. The entering flow was tripped by a fence and was passed over a $3$~m section of rough surface formed by road gravel to let the boundary layer developed before the flow had encountered the $5.15$~[m] long section of canopy. The leaf area index (LAI) of the model canopy is $0.47$. In other words, the wind-tunnel experiment represents a neutrally-stratified ABL flow over a homogeneous plant canopy in which the domain is $110h\times 38h\times 17h$. The estimated mean aerodynamic canopy height was $h=0.047$~m~\cite{Brunet94}. Given the mean velocity $U_h=2.88$~m/s at $z=h$, the Reynolds number $\mathcal Re = U_hh/\nu$ has a value of $9.0\times 10^3$ in this wind-tunnel study. %For a pedagogical reason, a sketch of the wind-tunnel setup has been reproduced in Fig~\ref{fig:fwt}$a$.

\subsection{Numerical setup}\label{sec:nstp}
In the present LES, we simulate a turbulent flow over a canopy of height $h=50$~[m] which is about $1\,000$ times larger than that of the wind-tunnel study mentioned above. The computational domain $1\,440\times 720\times 500$~[m$^3$] or $28.8h\times 14.4h\times 10h$~[m$^3$] of this study is the same as what was considered by the LES study of~\cite{Finnigan2009} (e.g. run A1). The domain is discretized into $256\times 128\times 123$ cells, where the mesh is uniform in both the horizontal directions (with $\Delta x = \Delta y = 5.625$~[m]). The vertical mesh is  stretched from $\Delta z=1.6$~[m] to $\Delta z=5.5$~[m], which is useful in capturing the mean shear near the bottom boundary. The flow is driven with a streamwise pressure gradient that is adjusted dynamically to yield a prescribed volume averaged  streamwise speed of $U_b$. The results with $U_b=4$~[m/s] yields a mean non-dimensional wind that agrees with the wind-tunnel measurements~\cite{Brunet94}. The Reynolds number ($U_bh/\nu$) for LES is $\mathcal Re=1.3\times 10^7$, which is $1\,477$ times larger than that in the wind tunnel experiment. Thus, the wind-tunnel data can be a scaled model of the flow represented by our LES. 

Since the drag coefficient $C_d$ decreases as $\mathcal Re$ increases~\cite{Brunet94}, a normalized form of the canopy drag force is considered to estimate $C_d$ for eq~(\ref{eq:LES1}), where $f_ih/U_b^2 = hC_d\mathcal A|\bm u|u_i$. We have observed that $hC_d\mathcal A=0.236$ yields a mean wind profile that agrees well with the wind profile measured in the wind-tunnel~\cite{Finnigan2009}. The plant area index (PAI), $h\mathcal A = 0.5$, yields $C_d=0.4725$ which is about $35$\% of the $C_d$ quoted by~\cite{Brunet94}. % Aiming at a reasonable comparison with the wind-tunnel measurements~\cite{Brunet94} and the reference LES data~\cite{Finnigan2009}, we keep $hC_da=0.236$ which is the same as what was reported by~\cite{Finnigan2009}.
The total simulation time for all the results is $6$~[h] unless it is mentioned otherwise. The data from the last $N$ time steps that account for a period of $3$~h are used for obtaining the average statistics, where $\bar u_i = (1/N)\sum_{k=1}^Nu_i^k$ and $\tau_{ij}^R = \overline{(u_i-\bar u_i)(u_j-\bar u_j)}$ are the resolved mean velocity and Reynolds stresses, respectively~\cite{Pope2000}. 

%First, we consider Monin-Obukhov similarity theory -- as depicted in Fig~\ref{fig:fwt}$a$. In the inertial sublayer of a neutrally stratified atmospheric boundary layer, the mean wind profile is approximately logarithmic. In the canopy sublayer and the roughness sublayer above the canopy, the log-law must be modified to reflect the region of modified turbulence and momentum flux. 
 
\begin{figure}
  \centering
  \begin{tabular}{cc}
    \includegraphics[height=6cm]{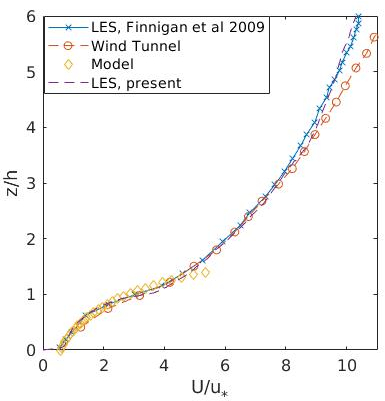}
    &\includegraphics[height=6cm]{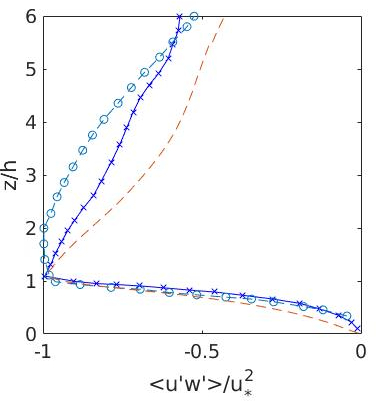}\\
    $(a)$ Mean $U(z)/u_*$ & $(b)$ Reynolds stress $\tau^R_{13}/u_*^2$
                                     
  \end{tabular}
  \caption{$(a)$ The result of the present LES ($--$) is compared with wind-tunnel data ($-o-$), a reference LES ($-\times-$), and a model profile of wind speed in the canopy layer $-\diamond-$. $(b)$ Vertical profile of $\tau^R_{13}/u_*^2$, present LES ($--$), wind-tunnel data ($-o-$), a reference LES ($-\times-$)}
  \label{fig:fwt}
\end{figure}

\subsection{Assessment of LES and wind-tunnel results}
We consider wind tunnel measurements to understand the dissipation of mean energy to the production of subgrid-scale energy. The subgrid-scale energy production by the mean wind $U(z)$ is the product of $\tau^R_{13} = \langle u'w'\rangle$ and $\partial U(z)/\partial z$. One expects that $\tau_{13}^R$ is likely to be negative if $U(z)$ is an increasing function of $z$. 
In our LES, the mean streamwise velocity $\bar u_1(x,y,z)$ was averaged over the entire horizontal plane at each vertical level to obtain $U(z) = \int_{x=0}^{28.8h}\int_{y=0}^{14.4h}\bar{u}_1(x,y,z)dydx$. The resulting vertical profile $U(z)/u_*$ of the normalized streamwise velocity, which is compared with the corresponding experimental data in Fig~\ref{fig:fwt}$a$. The wind profile is also compared with that of the LES of~\cite{Finnigan2009}. The results exhibit a good agreement among the three data sets. The deviation between results of LES and wind-tunnel data in the region of $z>h$ is primarily due to the differences in the outer boundary conditions~\cite{Finnigan2009}.

A critical difference between the atmospheric boundary layer flow of over a rough surface and that over a forest comes from the roughness sublayer. In Fig~\ref{fig:fwt}$a$, the vertical profile of the streamwise mean wind exhibits an inflection point at $z/h=1$, while the gradient of mean wind remains positive. Below the inflection point, the wind is well approximated by $e^{-\alpha(1-z/h)}$ with $\alpha=1.60$~\cite{Brunet94}. Such exponential change of wind is a fundamental difference of canopy flow with classical rough-wall turbulent boundary layer flows. Each of the experiment and LESs has a similar wind profile in the canopy region.

In Fig~\ref{fig:fwt}$b$, the stress $\tau_{13}^R$ is normalized by the friction velocity,  $u^2_* = \max\limits_{0\le z\le 10h}(-\tau_{13}^R)$. Clearly,  the vertical distribution of $\tau_{13}^R$ is negative, as expected. The most important finding is that the representation of SGS TKE by Q-criterion and vortex stretching in the SGS-w model is an acceptable representation of subgrid-scale turbulence in canopy flows. The guideline of studying SGS dissipation as a function of $Q$ and $\mathcal S\bm\omega$ comes from the earlier work of \cite{Dupont2008}, and \cite{Finnigan2009}. The role of vortex stretching in the SGS energy flux was thoroughly scrutinized by the LES study of rough-wall turbulent boundary layer~\cite{Chung2009}. While the study of coherent structures in canopy flows exists, to the best of our knowledge, this work is a first-time attempt in designing an SGS model for canopy flows, which is based on coherent structures.

%To validate this hypothesis, we consider the  vertical profile of the normalized mean streamwise velocity and the normalized Reynolds stress. %To compare the LES results with this wind-tunnel data, we have computed the mean vertical profile of the streamwise wind speed and the Reynolds stress, $\tau^R_{13}$.
%

It is worth discussing that we have not explicitly modeled the effects of wind shear. In other words, we have neither considered a damping function nor a boundary-layer specific blending of the length scale, $\Delta_{\hbox{\tiny LES}}$~(see \cite{Mason92}). The importance of the blending of length scale in boundary-layer meteorology in the context of the non-dynamic formulation of Deardorff's TKE model was studied by~\cite{Kurowski2018}. Recent findings of~\cite{Kurowski2018} also suggests for an equivalent dynamic approach, such as SGS-w model, for LES of canopy flows, which is further discussed below. %Designing the SGS-w model, we show how the SGS dissipation is dynamically adjusted to an asymptotic limit as the the ground is approached through a forest without using empirical adjustment of the length scale.

\section{Analysis and Discussion}\label{sec:and}
%
%\subsection{Wall-adaptive SGS model {\em vs} Lagrangian dynamic model and Deardorff's TKE model}
%
%To overcome the drawback of the classical Smagorinsky-Lilly model in canopy flows, the dynamic SGS modelling approaches for canopy flows are adopted by~\cite{Finnigan2009} (SGS-k) and \cite{Yan2017} (SGS-s). In those work, the value of $C_s$ is computed as the simulation progresses, usually by solving additional transport equations and the process involves additional test-filtering operations. Here, we explore the SGS-w closure for canopy flows, which does neither solve additional transport equations nor employs test-filtering operations. 
%%%To fully eliminate the barrier in the Lagrangian dynamic model and the Deardorff's TKE model, {\em i.e.} running these models without fully resolving the surface-layer (and canopy-layer), would require a major rework. To indicate that SGS-w model at least eliminates the need to solving additional transport equations in the following simulations, we evolve the velocity and pressure fields saved at the final time step of the LES run discussed in Section~\ref{sec:nstp}. The saved result is used as the initial condition for three new individual runs in which only the SGS closure is changed with respect to each other.  The results in the following sections indicate that the influences of stretched vortices~\cite[e.g.][]{Chung2009} can be modelled in canopy flows by engaging the rotation tensor $\mathcal R_{ij}$ into the eddy-viscosity of the SGS-w closure.
%
\subsection{Coherent structures and subgrid-scale motion in a canopy flow.}
\begin{figure}
  \centering
  \begin{tabular}{cc}
    %% \multicolumn{2}{c}{
    %% \includegraphics[height=4.5cm]{Q21600WALE}
    %% }\\
    %% \multicolumn{2}{c}{
    %% $(a)$ Q, SGS-w
    %% }\\
    \includegraphics[height=4.5cm]{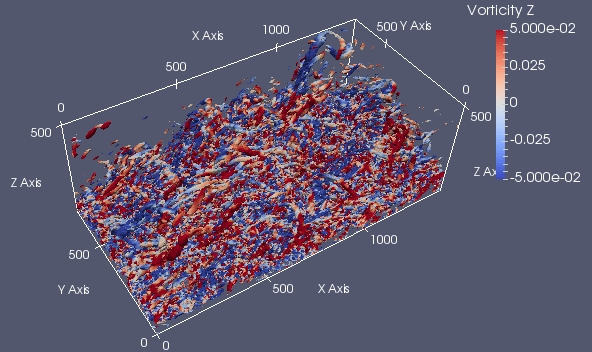} & \includegraphics[height=4.5cm]{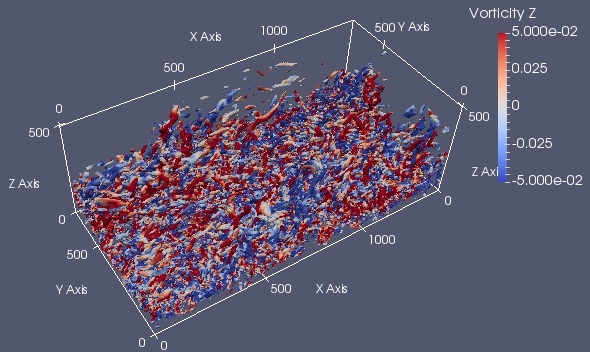}\\
    $(a)$ Q, SGS-w & $(b)$ Q, SGS-s \\
    \includegraphics[height=4.5cm]{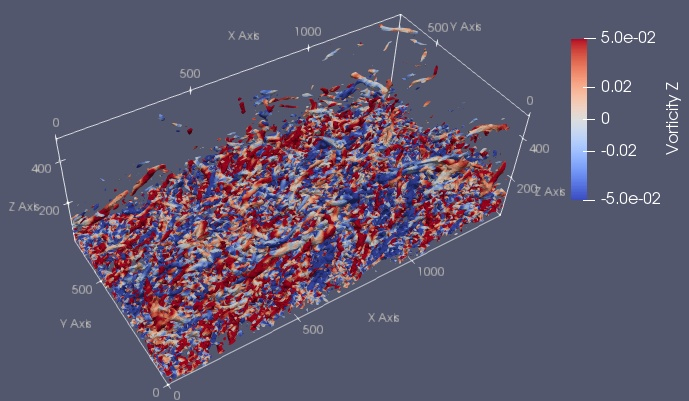}& \includegraphics[height=4.5cm,width=7.5cm]{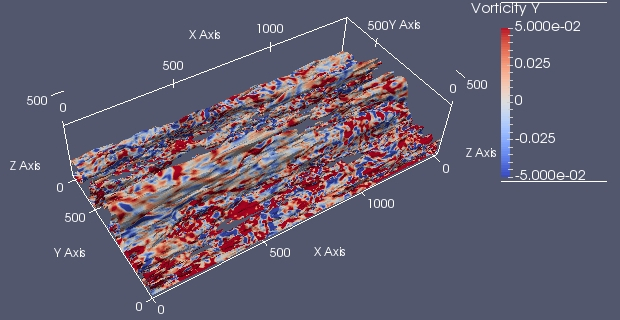}\\
    $(c)$ Q, SGS-d & $(d)$ $\tau^{R}_{13}$, SGS-w\\
  \end{tabular}
  \caption{A visualization of the coherent structures. Isosurfaces of the Q-criterion and the shear stress are colored by values of spanwise vorticity.  $(a)$ Q-criterion, shear-adjusted SGS-w model, $(b)$ Q-criterion, dynamic Lagrangian SGS-s model $(c)$ Q-criterion, Deardorff's TKE based SGS-d model, and $(d)$ shear-stress $\tau^R_{13}$, SGS-w model}
  \label{fig:Q}
\end{figure}

Consider the second invariant
\begin{equation}
  \label{eq:Q}
  \mathcal Q = \frac{1}{2}\left[\left(\frac{\partial u_i}{\partial x_j} - \frac{\partial u_j}{\partial x_i}\right)^2 - \left(\frac{\partial u_i}{\partial x_j} + \frac{\partial u_j}{\partial x_i}\right)^2\right]
\end{equation}
 of the velocity gradient tensor $\partial u_i/\partial x_j$, which can be interpreted as an indication  of the local pressure changes in the Navier-Stokes equations. A fluid region of $Q>0$ indicates that the rotation rate $\mathcal R_{ij} = \frac{\partial u_i}{\partial x_j} - \frac{\partial u_j}{\partial x_i}$ dominates over the strain rate $\mathcal S_{ij} = \frac{\partial u_i}{\partial x_j} + \frac{\partial u_j}{\partial x_i}$. However, $Q>0$ does not necessarily mean that the pressure is minimum within the region.
% 
% \begin{equation}
%   \label{eq:Rij}
%   \mathcal R_{ij}=\left(\frac{\partial u_i}{\partial x_j} - \frac{\partial u_j}{\partial x_i}\right)
% \end{equation}
%
Fig~\ref{fig:Q}($a,b,c$) demonstrates the isosurface of $Q=0.2U^2/h^2$ colored by the spanwise vorticity $\omega_y$ ({\em i.e.} $\mathcal R_{13}$), where the red and blue colors denote positive and negative values, respectively.  Note that a canopy flow contains the vorticity of the mean ABL flow $\langle U(z), 0, 0\rangle$, {\em i.e.} $\mathcal R_{13} = \partial u/\partial z-\partial w/\partial x$ that points toward the spanwise direction, as well as the vorticity associated with turbulent fluctuations, which is usually random. The spanwise rolls are likely to deform into inclined arched or hairpin-like structures -- the precise transition of which is flow-dependent~\cite{Bailey2016}. For a fully developed turbulent flow through a canopy, Fig~\ref{fig:Q} shows the the coherent structures while they are simultaneously advected, stretched, and tilted. One notes that the Reynolds shear-stress can be expressed in terms of velocity fluctuations in the direction of principal rate of strain, {\em i.e.} $\tau_{13}^R = (1/2)[\overline{(w'_*)^2} - \overline{(u'_*)^2}]$~(see the corresponding discussion in~\cite{Davidson2004}). Here, the notation~$(\cdot)'_*$ indicates that the co-ordinate system is rotated along the principal axes of the strain tensor. This means that in Fig~\ref{fig:Q} the negative (clock-wise) value of $\mathcal R_{13}$ is due to the Reynolds stress ({\em e.g.} Fig~\ref{fig:Q}$d$) associated with large fluctuations of $w'_*$ in the principal axes of the strain tensor. 
The qualitative description of the flow structures in Fig~\ref{fig:Q} also indicates that wind inside a forest canopy is linked to the morphology of canopy elements and  their aerodynamic response to airflow~\cite{Miri2017}.% In other words, present LES results complements the wind tunnel studies~\cite{Brunet94,Miri2017} through the visualization of three-dimensional flow structures. %Similar to mixing-layers and free shear flows, the inflection of the wind in a canopy flow is also connected to Kelvin-Helmholtz instabilities that are subsequently transformed into three-dimensional secondary instabilities. It is particularly because the nonlinear advection has the effect of passing energy from large to small scale ({\em i.e.} Richardson's picture of turbulence). The breakdown of coherent flow structures are responsible for the turbulent transfer of momentum (through sweeps and ejections) between the forest and the atmosphere. Although the turbulent transport by coherent structures is an imprecise statement to characterize the flow, it can be used to describe flow structures that are robust in the sense that they appear again and again in more or less in the same form. Nonetheless, these coherent flow structures are far from passive and they contribute significantly to the production of turbulent energy~\cite[see][]{Dupont2009}.
\subsection{Turbulence above a canopy layer}
\begin{figure}
  \centering
  \begin{tabular}{cc}
    \includegraphics[height=5cm]{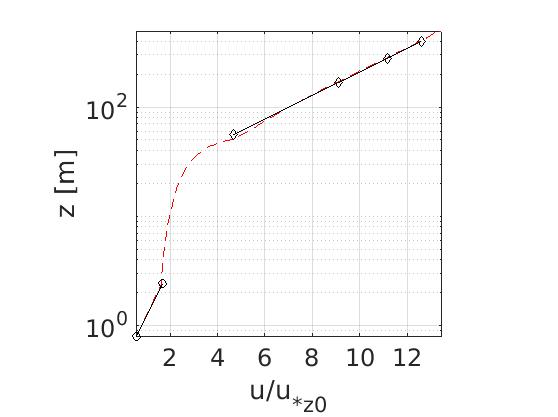}&    \includegraphics[height=5cm]{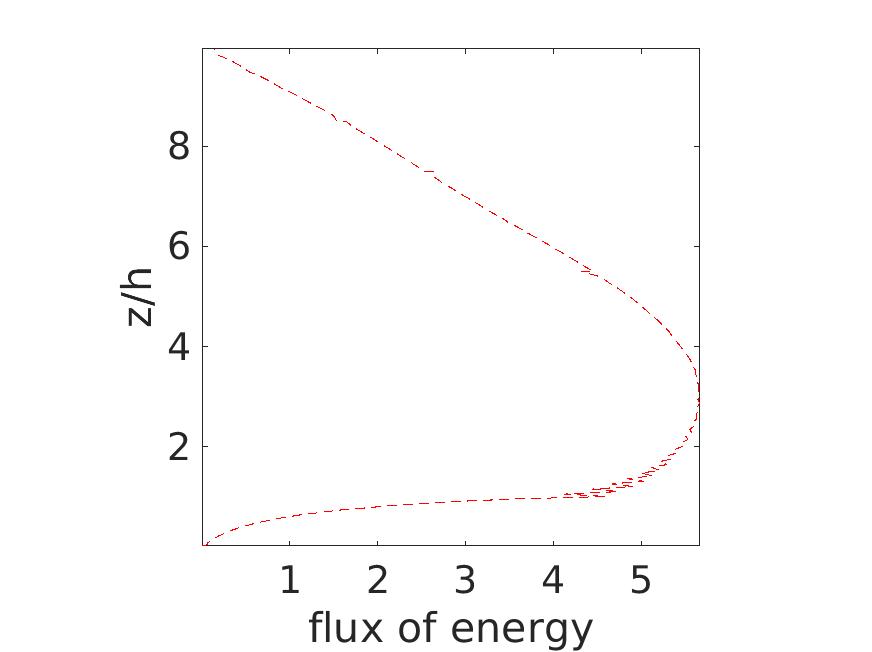}
  \end{tabular}
  \caption{$(a)$ The vertical profile of mean stream-wise flow showing two logarithmic region. One is above the canopy, which is characterized by the canopy friction velocity $u_{*_{z_1}}$.  The other is in the very bottom of the canopy, which is characterized by the surface friction velocity $u_{*_{z_0}}$. $(b)$ The energy flux $-\overline{u'w'}\partial u/\partial z$. Both plots are normalized by surface friction velocity $u_{*_{z_0}}$. }
  \label{fig:ent}
\end{figure}

 The structure of modified turbulence above the canopy layer have been studied most intensively~\cite{Finnigan2009}. The mean flow kinetic energy in the canopy layer is entrained from the free atmosphere, where kinetic energy is stored. In region above the canopy layer, the modified turbulence enhances a mechanism for the vertical mixing of the mean-flow kinetic energy and the momentum. From Eq~(\ref{eq:tke}) we estimate that the rate of dissipation per unit mass in the layer just above the canopy is $-\overline{u'w'}\partial u/\partial z = u^2_*(u_*/\kappa z)$, where the flow is restored to the neutrally stratified turbulent atmospheric boundary layer. Therefore, the amount of dissipation within the canopy layer $z_0 \le z\le z_1$ is given by $\int_{z_0}^{z_1}\epsilon(z)dz = (u^3_*/\kappa)\ln(z_1/z_0)$. Fig~\ref{fig:ent}$a$ demonstrate that two logarithmic regions exist in a fully developed turbulent flow through a canopy -- one is below the tree trunk and characterized by the surface friction velocity  $u_{*z_0}$. The other log-region is above the canopy edge, which is characterized by the canopy friction velocity is $u_{*z_1}$. The difference of the momentum fluxes between the two region, $u^2_{*z_1}-u^2_{*z_0}$, equals the momentum deficit caused by the canopy.

\begin{figure}
  \centering
  \begin{tabular}{ccc}
    \includegraphics[height=6cm]{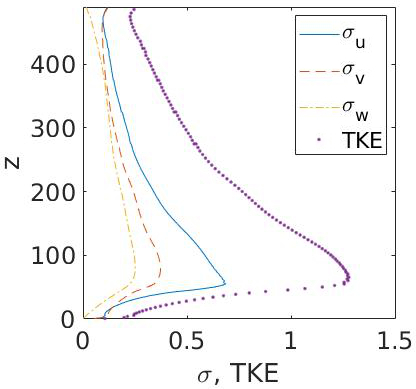}\\
    $(a)$ Scale-adaptive (SGS-w) \\
    \\
    \includegraphics[height=6cm]{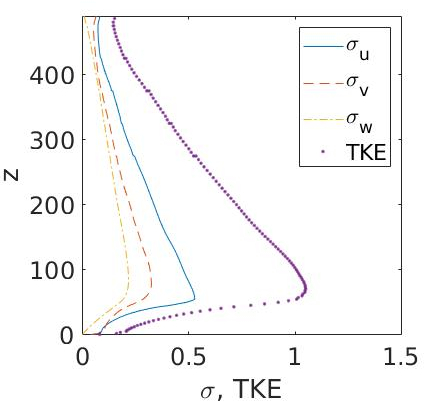} \\
    $(b)$ Deardorff's TKE (SGS-d)\\
    \\
    \includegraphics[height=6cm]{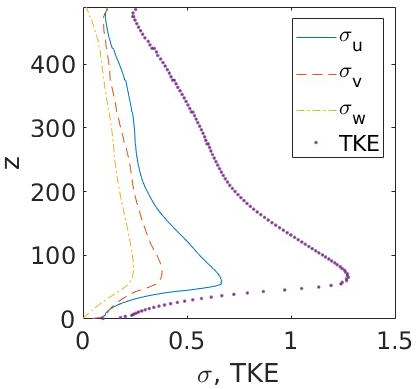}\\
    $(c)$ dynamic Lagrangian (SGS-s)\\
    %% \multicolumn{3}{c}{
    %% \includegraphics[height=5cm]{finniganDLAGsig}
    %% }\\
    %% \multicolumn{3}{c}{
    %% $(c)$ dynamic Lagrangian (SGS-s)
    %% }
  \end{tabular}
  \caption{Vertical profiles of the Reynolds stresses, $\sigma_u$, $\sigma_v$, $\sigma_w$, and TKE. $(a)$ Scale-adaptive algebraic eddy-viscosity model SGS-w, which does not solve any additional transport equation; $(b)$ Deardorff's TKE model SGS-d, which solves an additional transport equation, and $(c)$ Lagrangian dynamic model SGS-s, which solves two additional transport equations.}
  \label{fig:SGS}
\end{figure}

The statistical description of coherent structures in wall-bounded turbulent flows usually provides most of the phenomena exploited by turbulent flows over a canopy or another form of multiscale rough surfaces~\cite{Adrian2007}. In that view, the coherent structures depicted in Fig~\ref{fig:Q} are responsible to carry Reynolds stresses and to transport mean momentum. To illustrate the role that Reynolds stresses would play in the production and dissipation of TKE, the vertical distribution of the TKE is presented in Fig~\ref{fig:SGS}. The kinematic Reynolds stress, $\tau_{ij}^R = -\overline{ u'_iu'_j}$, is obtained by taking a time average of the solution during the last $3$ hours, treating this as an ensemble of large statistical samples distributed in the three-dimensional space. The TKE defined by $\tau_{ii}^R$ was averaged over the horizontal domain. Fig~\ref{fig:SGS}$a$ presents the vertical profiles of the diagonal components of the Reynolds stress, where $\sigma_u = \tau^R_{11}$, $\sigma_v = \tau^R_{22}$, and $\sigma_w = \tau^R_{33}$ are the variances of the velocity fluctuations, and half of their sum ({\em i.e.} TKE) represents turbulence intensity. The results are compared between three models, namely, SGS-w, SGS-d, and SGS-s. It is seen that the longitudinal velocity fluctuations are the largest because the shear production in a neutrally stable ABL initially feeds the energy into the $u$-component. The energy is subsequently distributed into the lateral components $v$ and $w$. As discussed earlier, the results of LES are expected to be relatively insensitive to the choice of the three SGS models if a majority of TKE is resolved. We see that each model predicts an overall expected behavior of the Reynolds stress. A variation of $18$\% in the prediction of turbulence intensity with respect to SGS-w and SGS-d suggest that the TKE carried by coherent structures are captured more directly by Eq~(\ref{eq:Nw}) than Eq~(\ref{eq:tke}). In the Lagrangian dynamic model, the time history of energy-carrying eddies are employed for adjusting the dissipation dynamically. A comparison of the result of locally-adaptive model (SGS-w) in Fig~\ref{fig:SGS}$a$ with that of Lagrangian dynamic model (SGS-s) in Fig~\ref{fig:SGS}$c$ also suggests that the consideration of $Q$ and $\mathcal S\bm\omega$ in the derivation of Eq~(\ref{eq:Nw}) help to capture the dynamical role of coherent structures without solving additional transport equations.

\begin{figure}
  \centering
  \begin{tabular}{cc}
    $(a)$ \\
    \includegraphics[height=7cm]{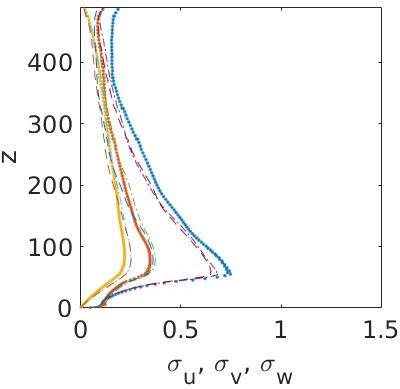}\\
    $(b)$ \\
    \includegraphics[height=7cm]{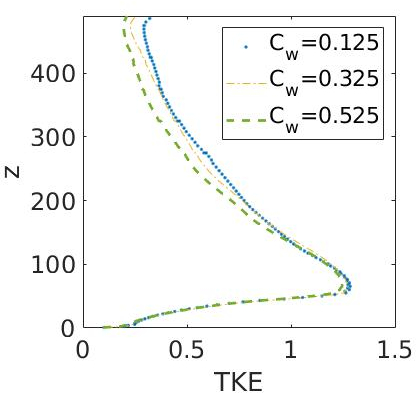}\\
  \end{tabular}
  \caption{The result of a sensitivity study of the model parameter $C_w$. In $(a)$ the legend is skipped for brevity, where $\cdots$ $C_w=0.125$, $-\cdot-$~$C_w=0.325$, and $--$ $C_w=0.525$. }
  \label{fig:CW}
\end{figure}

In Fig~\ref{fig:CW}, a brief sensitivity study of the model parameter $C_w$ of the SGS-w model is shown. Here, $C_w$ was estimated analytically by equating the eddy-viscosity given by Eq~(\ref{eq:Nw}) with the eddy-viscosity of Smagorinsky-Lilly model~(\ref{eq:nu}), which leads to $C_w^2 = C_s^2\Delta_{\hbox{\tiny LES}}\sqrt{2\mathcal S_{ij}\mathcal S_{ij}}/k_{\hbox{\tiny w}}^{1/2}$ (where $k^{1/2}_w$ is the expression within $[\cdot]$ in Eq~(\ref{eq:Nw}).) To estimate $C_w$, we considered a simulation of homogeneous isotropic turbulence in a periodic box  using $C_s=0.17$. It was found that a value of $C_w\sim 0.5$-$0.55$ persists for many eddy-turn over time for both decaying and forced turbulence. As depicted in Fig~\ref{fig:CW}, the vertical profiles of the resolved TKE for $C_w = 0.125,\,0.325,\,0.525$ are quite similar in the canopy sublayer. Based on our observation from several other numerical tests of the same canopy flow, we find that the modeled portion of the TKE $k_{\hbox{\tiny w}}$ is dynamically adjusted as the resolved flow varies by a change of $C_w$. Overall, a value of $C_w=0.325$ seems to be appropriate for canopy flows (see Fig~\ref{fig:CW}$b$).   


\subsection{Quadrant analysis}
\begin{figure}
  \centering
  \begin{tabular}{cc}
    $(a)$ & $(b)$ \\

    \includegraphics[height=4.5cm]{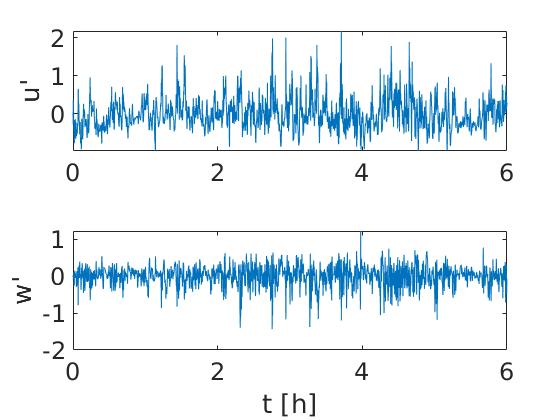}&
    \includegraphics[height=4.5cm]{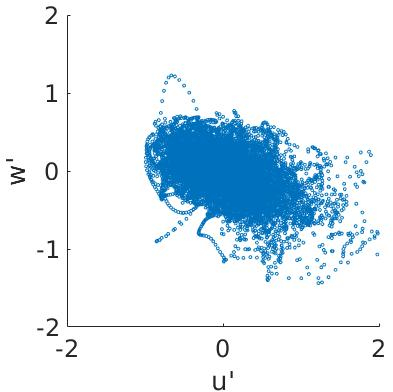} \\
     & \\
    $(c)$ & $(d)$ \\
    
    \includegraphics[height=4.5cm]{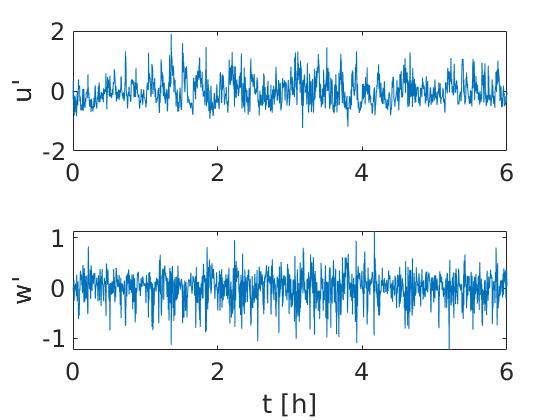}&
    \includegraphics[height=4.5cm]{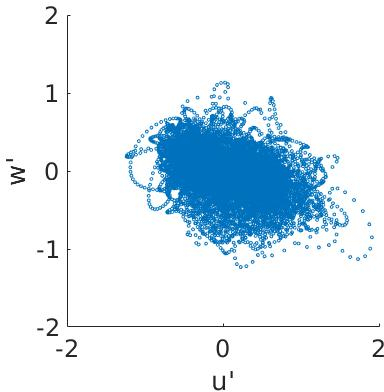} \\
     & \\
    
    $(e)$ & $(f)$ \\

    \includegraphics[height=4.5cm]{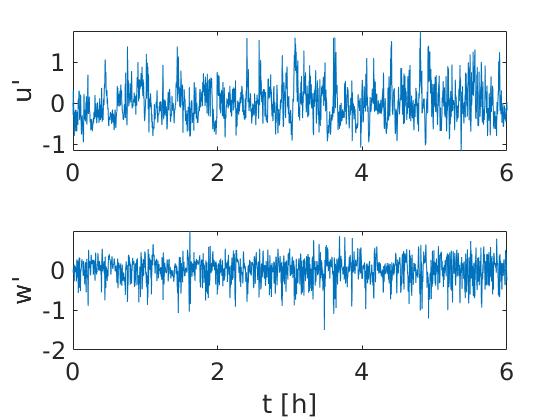}&
    \includegraphics[height=4.5cm]{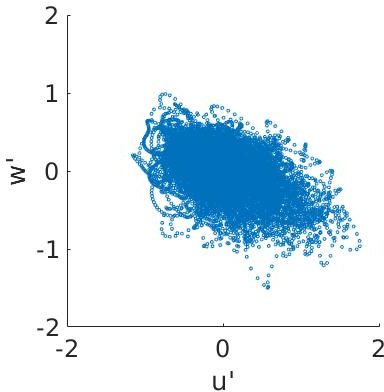} \\

    %% \includegraphics[height=3.5cm]{PDFuwDLAG}&
    %% \includegraphics[height=3.5cm]{PDFuwDKEQ}&
    %% \includegraphics[height=3.5cm]{PDFuwDLAG}\\
    %% $(g)$ & $(h)$ & $(i)$ \\
    %% SGS-w & SGS-d & SGS-s
  \end{tabular}
  \caption{Time series of fluctuations in the streamwise ($u'$) and vertical ($w'$) velocities. Left column: time series plots of $u'$ and $w'$ vs $t$; right column: scattered plot of $u'$ vs $w'$. $(a,b)$ wall-adaptive SGS-w model, $(c,d)$ Deardorff's TKE model, SGS-d, and $(e,f)$ dynamic Lagrangian SGS-w model}
  \label{fig:uwQ}
\end{figure}
%
\iffalse
\begin{figure}
  \centering
  \begin{tabular}{ccc}
    $(a)$ & $(b)$ & $(c)$ \\
    \includegraphics[height=3.5cm]{timeSeriesDKEQuw}&
    \includegraphics[height=3.5cm]{timeSeriesDKEQuw}&
                                                      \includegraphics[height=3.5cm]{timeSeriesDKEQuw}\\
    $(d)$ & $(e)$ & $(f)$ \\
    \includegraphics[height=3.5cm]{PDFuwDLAG}&
    \includegraphics[height=3.5cm]{PDFuwDLAG}&
    \includegraphics[height=3.5cm]{PDFuwDLAG}\\
    $(g)$ & $(h)$ & $(i)$ \\
    \includegraphics[height=4.5cm]{scatterWALEuw} &
    \includegraphics[height=4.5cm]{scatterDKEQuw} &
    \includegraphics[height=4.5cm]{scatterDLAGuw} \\
    SGS-w & SGS-d & SGS-s
  \end{tabular}
  \caption{Time variation of velocity fluctuation in the streamwise ($u'$) and vertical ($w'$) directions. top row: scattered plot of $u'$ vs $w'$; middle row: time series plots of $u'$ and $w'$, and bottom row: Probability Density Function of $u'$ and $w'$. left column: wall-adaptive SGS-w model, middle column: Deardorff's TKE model, SGS-d, and right column: dynamic Lagrangian SGS-w model}
  \label{fig:uwQ}
\end{figure}
\fi
%
While a flow visualization technique can identify coherent structures, the quadrant analysis technique is a powerful yet simple method, to investigate the contributions of the coherent motions. The instantaneous stress $-\overline{u'w'}$ is viewed in the four quadrants of $u'-w'$ plane, namely, Q1~($u'>0,w'>0$), Q2~($u'<0,w'>0$), Q3~($u'<0,w'<0$), and Q4~($u'>0,w'<0$). In the context of ABL studies this method holds a lot of promise for a single level sonic anemometer data to determine the largest possible contributions to $-\overline{u'w'}$, such as statistical properties of strong ``ejection-like'' ($u'<0,\,w'>0$,\, Q2) or ``sweep-like'' ($u'>0,\,w'<0$,\, Q4) bursting phenomena of boundary layer turbulence. Ref~\cite{Zhou99} extracted coherent structures from the DNS of a channel flow by linear stochastic estimation of ``ejection-like'' near-wall events. Using Haar wavelet transform, ref~\cite{Watanabe2004} observed that sweep-ejection cycle has a dominant contribution to the Reynolds stress. Ref~\cite{Finnigan2009} observed that the conjunction of Q2 and Q4 events produces the location of the coherent scalar microfront, and that the sweep-ejection cycle is also associated with the breaking of symmetry in flows over a vegetation canopy.

For the purpose of understanding the predictions of SGS motion by the SGS-w model, Eq~(\ref{eq:Nw}), Fig~\ref{fig:uwQ} illustrates a qualitative comparison of sweep- and ejection-like events among three SGS models. The time series of the resolved SGS velocities $u'$ (=$u - \bar u$) and $w'$ (=$w-\bar w$) are displayed in Fig~\ref{fig:uwQ}$(a,c,e)$, where $u$ and $w$ are resolved velocity computed at a height of $z/h=1$ on the vertical centerline of the computational domain. The nature of the time series of the velocity fluctuations with respect to three SGS models is similar, except the intermittent burst of horizontal velocity fluctuations are relatively protuberant in SGS-w/s models than the SGS-d models. This is consistent with the similarity of the flow structures depicted in Fig~\ref{fig:SGS}. It is evident implicitly that the rate ($\tau^R_{13}\mathcal S_{13}$) of mixing by the Reynolds stress in the forest edge -- due to passing of energy from the mean flow to the turbulence -- is enhanced by resolved shear $\mathcal S_{13}$.  Fig~\ref{fig:uwQ}$(b,d,f)$ demonstrate the scattered plots of the velocity fluctuations on the $u'$-$w'$ plane. Briefly, the ejection (Q2) of low-momentum are associated with the sweeps (Q4) of high momentum fluids. It is worth mentioning that an overall similarity among the plots in  Fig~\ref{fig:uwQ}$(b,d,f)$ indicates the similar nature of three dynamic modeling approaches.

\section{Conclusion}
In this article, a numerical study of the energy transfer between the resolved- and the subgrid-scale motions in a turbulent flow over a forest-like canopy is summarized. LES and wind-tunnel measurements of the mean vertical wind profile and the Reynolds stress suggests that the subgrid-scale energy transfer is strongly correlated with the coherent structures and vortex stretching mechanism. This result is consistent with the previous findings of \cite{Dupont2008} and \cite{Finnigan2009}.

In  particular,  we find that the advancement of the scale-adaptive large-eddy simulation framework is a potential methodology that correlates the SGS dissipation of TKE to the vortex stretching mechanism through the second invariant of the square of the velocity gradient tensor. Considering the form-drag experienced by a forest-like canopy in the form of three-dimensional Forchheimer stress, the wind profiles simulated by our LES agree with the corresponding profiles of wind-tunnel measurements. Incorporating the viscous drag in the form of Darcy-Forcheimer stress, this work observes two log-regions: one below the tree trunk and one above the canopy edge. The bottom log-region is characterized by the friction of the surface, and the upper log-region is characterized by the friction of the canopy with the air-flow aloft.

It is worth mentioning that this study considers three strategies of the dynamic modeling approach for canopy flows. We have analyzed scale-adaptive large-eddy simulation for canopy flows, where neither the canopy elements nor the viscous sublayer is fully resolved. The results show that capturing the effects of coherent structures captured about $18$\% more SGS TKE with respect to resolving the dynamics of SGS TKE through a transport equation.

There is a substantial investigation of canopy flows, for example, see the work of~\cite{Raupach96} and \cite{Finnigan2009}. However, there remains several questions in the context of SGS turbulence modeling. As nature of the canopy flow shares between mixing layer and turbulent boundary layer, it is an ideal candidate for understanding the role of self-amplification of the strain-rate with that of vortex stretching. We think that such consideration may help our future studies on understanding the transition of scales in atmospheric turbulence, particularly in `grey-zone' turbulence.

\section*{Acknowledgments}
The authors acknowledge the computational facilities provided by the Compute Canada (graham.computecanada.ca). JMA acknowledges financial support from NSERC, and MASB acknowledges President awards and SWASP funds of the Memorial University of Newfoundland.

%\begin{acknowledgements}
%\end{acknowledgements}


\bibliographystyle{spmpsci}      % mathematics and physical sciences
\bibliography{forestCanopy}   % name your BibTeX data base

\begin{thebibliography}{10}
\providecommand{\url}[1]{{#1}}
\providecommand{\urlprefix}{URL }
\expandafter\ifx\csname urlstyle\endcsname\relax
  \providecommand{\doi}[1]{DOI~\discretionary{}{}{}#1}\else
  \providecommand{\doi}{DOI~\discretionary{}{}{}\begingroup
  \urlstyle{rm}\Url}\fi

\bibitem{Adrian2007}
Adrian, R.J.: Hairpin vortex organization in wall turbulence.
\newblock Physics of Fluids \textbf{19}(4), 041301 (2007)

\bibitem{Alam2015}
Alam, J., Islam, M.R.: A multiscale eddy simulation methodology for the
  atmospheric ekman boundary layer.
\newblock Geophysical \& Astrophysical Fluid Dynamics \textbf{109}(1), 1--20
  (2015)

\bibitem{Alam2018}
Alam, J.M., Fitzpatrick, L.P.J.: Large eddy simulation of urban boundary layer
  flows using a canopy stress method.
\newblock Computers \& Fluids \textbf{171}, 65--78 (2018)

\bibitem{Chow2019}
Arthur, R.S., Mirocha, J.D., Lundquist, K.A., Street, R.L.: Using a canopy
  model framework to improve large-eddy simulations of the neutral atmospheric
  boundary layer in the weather research and forecasting model.
\newblock Monthly Weather Review \textbf{147}(1), 31--52 (2019)

\bibitem{Bailey2016}
Bailey, B.N., Stoll, R.: The creation and evolution of coherent structures in
  plant canopy flows and their role in turbulent transport.
\newblock Journal of Fluid Mechanics \textbf{789}, 425–460 (2016)

\bibitem{Belcher2012}
Belcher, S.E., Harman, I.N., Finnigan, J.J.: The wind in the willows: Flows in
  forest canopies in complex terrain.
\newblock Annual Review of Fluid Mechanics \textbf{44}(1), 479--504 (2012)

\bibitem{Brunet94}
Brunet, Y., Finnigan, J., Raupach, M.: A wind tunnel study of air flow in
  waving wheat: single-point velocity statistics.
\newblock Boundary-Layer Meteorology \textbf{70}(1-2), 95--132 (1994)

\bibitem{Chung2009}
Chung, D., Pullin, D.I.: Large-eddy simulation and wall modelling of turbulent
  channel flow.
\newblock Journal of Fluid Mechanics \textbf{631}, 281–309 (2009)

\bibitem{Davidson2004}
Davidson, P.A.: Turbulence - An Introduction for Scientists and Engineers.
\newblock Oxford University Press (2004)

\bibitem{DeLemos2006}
De~Lemos, M.: Turbulence in Porous Media: Modeling And Applications (2006)

\bibitem{Deardorff70}
Deardorff, J.W.: A numerical study of three-dimensional turbulent channel flow
  at large reynolds numbers.
\newblock J. Fluid Mech. \textbf{41(2)}, 453--480 (1970)

\bibitem{Deardorff73}
Deardorff, J.W.: The use of subgrid transport equations in a three-dimensional
  model of atmospheric turbulence.
\newblock Journal of Fluids Engineering \textbf{95}, 429 (1973)

\bibitem{Deardorff74}
Deardorff, J.W.: Three-dimensional numerical study of the height and mean
  structure of a heated planetary boundary layer.
\newblock Boundary-Layer Meteorology \textbf{7}(1), 81--106 (1974)

\bibitem{Deardorff80}
Deardorff, J.W.: Stratocumulus-capped mixed layers derived from a
  three-dimensional model.
\newblock Boundary-Layer Meteorology \textbf{18}(4), 495--527 (1980)

\bibitem{Dupont2008}
Dupont, S., Brunet, Y.: Edge flow and canopy structure: A large-eddy simulation
  study.
\newblock Boundary-Layer Meteorology \textbf{126}(1), 51--71 (2008)

\bibitem{Dupont2009}
Dupont, S., Brunet, Y.: Coherent structures in canopy edge flow: a large-eddy
  simulation study.
\newblock Journal of Fluid Mechanics \textbf{630}, 93–128 (2009)

\bibitem{Dwyer97}
Dwyer, M.J., Patton, E.G., Shaw, R.H.: Turbulent kinetic energy budgets from a
  large-eddy simulation of airflow above and within a forest canopy.
\newblock Boundary-Layer Meteorology \textbf{84}(1), 23--43 (1997)

\bibitem{Finnigan2000}
Finnigan, J.: Turbulence in plant canopies.
\newblock Annual Review of Fluid Mechanics \textbf{32}(1), 519--571 (2000)

\bibitem{Finnigan2009}
Finnigan, J.J., Shaw, R.H., Patton, E.G.: Turbulence structure above a
  vegetation canopy.
\newblock Journal of Fluid Mechanics \textbf{637}, 387–424 (2009)

\bibitem{Germano91}
Germano, M., Piomelli, U., Moin, P., Cabot, W.H.: A dynamic subgrid scale eddy
  viscosity model.
\newblock Physics of Fluids A \textbf{3}(7), 1760--1775 (1991)

\bibitem{Konstantin2018}
Kr{\"o}niger, K., Banerjee, T., Roo, F.D., Mauder, M.: Flow adjustment inside
  homogeneous canopies after a leading edge -- {A}n analytical approach backed
  by les.
\newblock Agricultural and Forest Meteorology \textbf{255}, 17 -- 30 (2018)

\bibitem{Kurowski2018}
Kurowski, M.J., Teixeira, J.: A scale-adaptive turbulent kinetic energy closure
  for the dry convective boundary layer.
\newblock Journal of the Atmospheric Sciences \textbf{75}(2), 675--690 (2018)

\bibitem{Leveque2007}
LÉVÊQUE, E., TOSCHI, F., SHAO, L., BERTOGLIO, J.P.: Shear-improved
  smagorinsky model for large-eddy simulation of wall-bounded turbulent flows.
\newblock Journal of Fluid Mechanics \textbf{570}, 491–502 (2007)

\bibitem{Marjoribanks2016}
Marjoribanks, T., Hardy, R., Lane, S., Parsons, D.: Does the canopy mixing
  layer model apply to highly flexible aquatic vegetation? insights from
  numerical modelling.
\newblock Environmental Fluid Mechanics \textbf{17} (2016)

\bibitem{Mason92}
Mason, P.J., Thomson, D.J.: Stochastic backscatter in large-eddy simulations of
  boundary layers.
\newblock Journal of Fluid Mechanics \textbf{242}, 51–78 (1992)

\bibitem{Meneveau96}
Meneveau, C., Lund, T.S., Cabot, W.H.: A lagrangian dynamic subgrid-scale model
  of turbulence.
\newblock Journal of Fluid Mechanics \textbf{319}, 353–385 (1996).
\newblock \doi{10.1017/S0022112096007379}

\bibitem{Miri2017}
Miri, A., Dragovich, D., Dong, Z.: Vegetation morphologic and aerodynamic
  characteristics reduce aeolian erosion.
\newblock Scientific Reports \textbf{7}(1), 12831 (2017)

\bibitem{Moeng84}
Moeng, C.H.: A large-eddy-simulation model for the study of planetary
  boundary-layer turbulence.
\newblock Journal of the Atmospheric Sciences \textbf{41}(13), 2052--2062
  (1984)

\bibitem{Moeng2015}
Moeng, C.H., Sullivan, P.: Encyclopedia of Atmospheric Sciences, 2nd Edition,
  vol.~4, pp. 232--240.
\newblock Elsevier Ltd, Academic Press (2015)

\bibitem{Nicoud99}
Nicoud, F., Ducros, F.: Subgrid-scale stress modelling based on the square of
  the velocity gradient tensor.
\newblock Flow, Turbulence and Combustion \textbf{62}(3), 183--200 (1999)

\bibitem{Nicoud2011}
Nicoud, F., Toda, H.B., Cabrit, O., Bose, S., Lee, J.: Using singular values to
  build a subgrid-scale model for large eddy simulations.
\newblock Physics of Fluids \textbf{23}(8), 085106 (2011)

\bibitem{Park2015}
Park, S.B., Baik, J.J., Han, B.S.: Large-eddy simulation of turbulent flow in a
  densely built-up urban area.
\newblock Environmental Fluid Mechanics \textbf{15}(2), 235--250 (2015)

\bibitem{Philips2013}
Philips, D.A., Rossi, R., Iaccarino, G.: Large-eddy simulation of passive
  scalar dispersion in an urban-like canopy.
\newblock Journal of Fluid Mechanics \textbf{723}, 404–428 (2013)

\bibitem{Pope2000}
Pope, S.B.: {Turbulent Flows}.
\newblock Cambridge University Press (2000)

\bibitem{Raupach96}
Raupach, M.R., Finnigan, J.J., Brunei, Y.: Coherent eddies and turbulence in
  vegetation canopies: The mixing-layer analogy.
\newblock Boundary-Layer Meteorology \textbf{78}(3), 351--382 (1996)

\bibitem{Shaw92}
Shaw, R.H., Schumann, U.: {Large-Eddy Simulation of Turbulent Flow above and
  within a Forest}.
\newblock Boundary Layer Meteorology \textbf{61}, 47--64 (1992)

\bibitem{smagorinsky}
Smagorinsky, J.: General circulation experiments with the primitive equations.
\newblock Monthly Weather Review \textbf{91}, 99 (1963)

\bibitem{Stvredova2012}
St{\v{r}}edov{\'a}, H., Podhr{\'a}zsk{\'a}, J., Litschmann, T., St{\v{r}}eda,
  T., Ro{\v{z}}novsk{\`y}, J.: Aerodynamic parameters of windbreak based on its
  optical porosity.
\newblock Contributions to Geophysics and Geodesy \textbf{42}(3), 213--226
  (2012)

\bibitem{Sullivan94}
Sullivan, P.P., McWilliams, J.C., Moeng, C.H.: A subgrid-scale model for
  large-eddy simulation of planetary boundary-layer flows.
\newblock Boundary-Layer Meteorology \textbf{71}(3), 247--276 (1994)

\bibitem{Taylor32}
Taylor, G.I.: The transport of vorticity and heat through fluids in turbulent
  motion.
\newblock Proceedings of the Royal Society of London. Series A, Containing
  Papers of a Mathematical and Physical Character \textbf{135}(828), 685--702
  (1932)

\bibitem{Trias2015}
Trias, F.X., Folch, D., Gorobets, A., Oliva, A.: Building proper invariants for
  eddy-viscosity subgrid-scale models.
\newblock Physics of Fluids \textbf{27}(6), 065103 (2015)

\bibitem{Watanabe2004}
Watanabe, T.: Large-eddy simulation of coherent turbulence structures
  associated with scalar ramps over plant canopies.
\newblock Boundary-Layer Meteorology \textbf{112}(2), 307--341 (2004)

\bibitem{Yan2017}
Yan, C., Huang, W.X., Miao, S.G., Cui, G.X., Zhang, Z.S.: Large-eddy simulation
  of flow over a vegetation-like canopy modelled as arrays of bluff-body
  elements.
\newblock Boundary-Layer Meteorology  (2017)

\bibitem{Zhou99}
Zhou, J., Adrian, R.J., Balachandar, S., Kendall, T.M.: Mechanisms for
  generating coherent packets of hairpin vortices in channel flow.
\newblock Journal of Fluid Mechanics \textbf{387}, 353–396 (1999).
\newblock \doi{10.1017/S002211209900467X}

\end{thebibliography}

%\begin{thebibliography}{}
% etc
%\end{thebibliography}

\end{document}